\documentclass[fleqn,12pt,twoside]{article}
\usepackage{espcrc1}

\usepackage{graphicx}

\title{Precise laser spectroscopy  of the antiprotonic helium atom
and CPT test on antiproton mass and charge}

\author{H. Yamaguchi\address[UTH]{Department of Physics, University of Tokyo,\\ 
        7-3-1 Hongo, Bunkyo-ku, Tokyo 113-0033, Japan},
        J. Eades\address[CERN]{CERN, \\
        CH-1211 Geneva 23, Switzerland},
        R. S. Hayano\addressmark[UTH], 
        M. Hori\addressmark[CERN],
        D. Horv\'{a}th\address[KFKI]{KFKI Research Institute For Particle 
	and Nuclear Physics, \\
	H-1525 Budapest, Hungary}
	T. Ishikawa\addressmark[UTH], 	
	 B. Juh\'{a}sz\address[INR]{Institute of Nuclear Research 
	of the Hungarian Academy of Sciences, \\
	H-4001 Debrecen, Hungary}
	J. Sakaguchi\addressmark[UTH], 
	H. A. Torii\address[UTK]{Institute of Physics, University of Tokyo, \\
	Komaba, Meguro-ku, Tokyo 153-8902, Japan} 
	E. Widmann\addressmark[UTH],
        and
        T. Yamazaki\address[RIKENRI]{RI Beam Science Laboratory, RIKEN, \\
	Wako, Saitama 351-0198, Japan}
	}
\begin{document}

\maketitle

\begin{abstract}
We have measured twelve transition frequencies of 
the antiprotonic helium atom ($\bar{p}$He$^+$) with precisions of 0.1--0.2 ppm
using a laser spectroscopic method.
The agreement between the experiment and theories was so good 
that we can put a limit on the proton-antiproton mass (or charge) 
difference.  The new limit is expected to be much smaller
than the already published value, 60 ppb ($6 \times 10^{-8}$).

\end{abstract}

\section{Introduction}

The antiprotonic helium atom ($\bar{p}$He$^+$) is a
 three-body system consisting of an antiproton, an electron,
and a helium nucleus.
Some states (quantum numbers; $n$, $l$ $\sim$ 38) 
of this exotic atom are known to live 
anomalously long (lifetime $\sim$ 3 $\mu$s)
for a system including an antiproton. 
Since its discovery in 1991 \cite{Iwasaki1991,Yamazaki1993},
the nature of this antiprotonic atom has been studied extensively, and 
precise measurements of its enegy levels have been carried out 
using a laser spectroscopic method.

In the last few years we have performed sub-ppm
laser spectroscopy on many transitions of 
the antiprotonic atom at CERN AD (Antiproton Decelerator) \cite{Hori2001}.
We have done a CPT test on proton-antiproton mass and charge 
differences by comparing the experiment with theories,
as the theories use the known proton mass value
for the antiproton mass.

\section{Experimental Setup}

Our current experimental scheme is in principle the same as
the one described in Refs.~\cite{Hori2001,HoriNIMA}.
A major improvement is that we started to use a new radiofrequency quadrupole
decelerator (RFQD) \cite{Pirkl}. 
The apparatus can decelerate antiprotons from 5.3 MeV to below 100 keV
by a RFQ electric field. 
Previously, we had to use a gas target dense enough to stop 5.3 MeV 
antiprotons (atomic density $\rho \sim 10^{21}$ cm$^{-3}$),
but such a high density can shift the resonant frequency significantly.
By using the RFQD, antiprotons can be stopped in a far lower density target
($\rho =  10^{17}$--$10^{18}$ cm$^{-3}$),
thus we can directly measure the transition frequencies at ``zero-density'', 
instead of taking several scans at different target densities 
and making an extrapolation.


\section{Analysis}

\subsection{Transition frequency}

The analysis method for the transition frequencies
is also concisely described in the Refs.~\cite{Hori2001,HoriNIMA}.
Here we summarize the three procedures
used for the deduction of the frequencies.

\begin{itemize} 
\item Fitting the frequency profiles of the resonances \\
Each resonance profile was
fitted with a sum of two identical-shaped Voigt functions 
separated by the theoretical hyperfine splitting \cite{Widmann1997,BK1998}. 
The uncertainty of the fitting is $\Delta \nu = $ 20--60 MHz,
which mainly comes from statistical fluctuations,
instability of the laser power,
and irregular frequency distributions of the laser pulse.

\item Absolute frequency calibration \\
Our wavelength meter is absolutely calibrated against 
 molecular ro-vibrational lines of iodine  \cite{iodine1,iodine2} or
tellurium, or atomic lines of neon or argon by optogalvanic spectroscopy.
The accuracy of the calibration is 20--50 MHz.

\item Density extrapolaration \\
For high density (non-RFQD) scans, transition frequencies at zero-density 
were obtained by linear extrapolation of the several central frequencies 
at different densities.
Due to the limited accuracy of the temperature and pressure measurement, 
this procedure causes 20--50 MHz of uncertainty.

\end{itemize}
By taking into account all of the above, 
the overall precision $\Delta \nu / \nu$ was 0.1--0.2 ppm.

\subsection{CPT test}
The agreement between the experiment and theories enables 
us to put a limit on the proton-antiproton mass (or charge) difference,
which is equal to zero if the CPT theorem holds.
Here we describe how we could obtain the CPT limit 
from the experimental results.

The charge-to-mass ratio of antiproton $Q_{\bar{p}}/M_{\bar{p}}$ 
has been measured to a high precision of 
$9 \times 10^{-11}$ by a Penning trap experiment \cite{Gab1999}.
Our experiment measures transition frequencies, which are 
differences of energy levels.
The frequency of an antiprotonic transition should depend
on the antiproton mass  and 
the charge. 
If we assume a hydrogen-like system, this dependence 
can be written as 
\begin{equation}
\hbar\nu = \Delta E \propto M_{\bar{p}}Q_{\bar{p}}^2. 
\end{equation}

However, the actual dependence is different from the above,
 because the system consists of three particles 
and the antiproton is quite heavy.
Instead, we can write the realistic dependence by introducing a parameter $f$,
which depends on the transition:
\begin{equation}
\Delta E \propto M_{\bar{p}}Q_{\bar{p}}^{f-1} = 
({Q_{\bar{p}}}/{M_{\bar{p}}})^{-1} Q_{\bar{p}}^{f} = 
({Q_{\bar{p}}}/{M_{\bar{p}}})^{f-1} M_{\bar{p}}^{f}.
\end{equation}
If we fix the well-known charge-to-mass ratio,
$\Delta E$ is simply proportional to $Q_{\bar{p}}^f$ and  $M_{\bar{p}}^f$.
The parameter $f = $ 2--6
 was theoretically evaluated by Kino \cite{Kinopr}
by calculating the shift of the transition frequency 
when the input proton mass value is changed slightly.

Then, the CPT limit parameter $\delta$ can be obtained by 
comparing the experimental and theoretical values:
\begin{equation}
\delta = 
\frac{Q_p + Q_{\bar{p}}}{Q_p} =
\frac{M_p - M_{\bar{p}}}{M_p} =
\frac{1}{f} \frac{\nu_{th}- \nu_{exp}}{\nu_{exp}},
\end{equation}
since the theories use the precisely-known proton mass as the 
antiproton mass in the calculations.
The limits of all the twelve transitions 
were averaged to obtain the final result.

\section{Results}
\begin{figure}[htb]
\caption{Precise comparison of the experimental and the theoretical transition frequencies of $\bar{p}$$^4$He$^+$  \cite{Hori2001}.} 
\label{fig:comparison2000}
\includegraphics{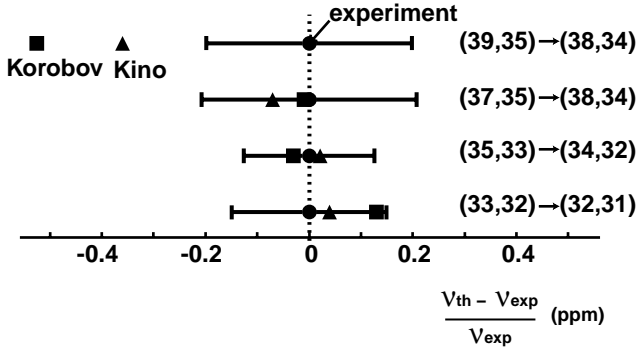}
\end{figure}

Fig.~\ref{fig:comparison2000} shows the comparison of the experimental and the theoretical transition frequencies published in 2001 \cite{Hori2001}.
In this figure, the experimental values are centered on the 
dotted lines and their precisions $\Delta\nu_{exp}/\nu_{exp}$ are shown as 
error bars.
 Two independent calculations by Korobov \cite{Korobov1997} and Kino \cite{Kino1999} are compared with our results.
Although not shown in this figure, eight more states of 
$\bar{p}^4$He$^+$ and $\bar{p}^3$He$^+$ measured in 2001--2002 
are in analysis.

\clearpage

Table~\ref{table:res_summary} is the summary of our progress.
From the measurements in 2000, we obtained a CPT limit of 60 ppb with a confidence level of 90\% 
\cite{Hori2001}. 
Since then, many transitions including those of $\bar{p}^3$He$^+$   
were measured, and the RFQD contributed to an efficient data-taking.
As a result, the CPT limit is, according to   
our preliminary analysis so far,
greatly improved by a factor 3 or better.
The final result of the analysis will be presented in \cite{Hori2003}.

\begin{table}[htb]
\caption{Summary of the progress of our measurement at AD.}
\label{table:res_summary}
\begin{tabular}{p{150pt}|p{120pt}|p{120pt}}
\hline
Year           & 2000 & 2001--2002 \\
\hline
Precision of each frequency & 0.2 ppm & 0.1 -- 0.2 ppm \\
Measured transitions & 4 & 12 \\
$\bar{p}$He isotope               & $\bar{p}$$^4$He only  & $\bar{p}$$^3$He, $\bar{p}$$^4$He \\
RFQD           & Not used & Used \\
CPT limit $|\delta|$ (90\% C.L.)   & 60 ppb & $<$ 20 ppb (preliminary)  \\
Publication & \cite{Hori2001}, referred by \cite{PDG} & \cite{Hori2003}(to be published) \\
\hline
\end{tabular}\\
\end{table}

We are grateful to CERN PS division for their help,
V.I. Korobov, D.D. Bakalov and Y. Kino for useful discussions.
This work was supported by the Grant-in-Aid for Creative Basic Research
(Grant No. 10NP0101) of Monbukagakusho of Japan,
and the Hungarian Scientific Research Fund (Grant Nos. OTKA T033079
and TeT-Jap-4/00).

\end{document}